%% file: GES-approach-arXiv-v2.tex
\documentclass[smallextended]{svjour3} 
\usepackage{bm,bbm}
\usepackage{times}
\usepackage{graphicx} 
\usepackage{subfigure,color}
\usepackage[draft]{fixme}
\usepackage{changes,color}
\usepackage[super]{nth}
\usepackage{amsmath,amsfonts}
\definecolor{myurlcolor}{rgb}{0,0,0.7}
\definecolor{myrefcolor}{rgb}{0.8,0,0}
\usepackage{hyperref}
\hypersetup{colorlinks, linkcolor=myrefcolor,
citecolor=myurlcolor, urlcolor=myurlcolor}
\usepackage{hyperref}
\usepackage{mathtools}

\DeclarePairedDelimiter\floor{\lfloor}{\rfloor}
\input{komendy_osid.tex}

\newtheorem{thm}{Theorem}

\newtheorem{lem}[thm]{Lemma}
\newtheorem{defi}[thm]{Definition}

\newtheorem{fakt}[thm]{Fact}
\newcommand{\beu}{\begin{equation}}
\newcommand{\eeu}{\end{equation}}
\newcommand{\be}{\begin{eqnarray}}
\newcommand{\ee}{\end{eqnarray}}

\newcommand{\ce}{\mathbb{C}}
\newcommand{\cee}[1]{\mathbb{C}^{#1}}

\begin{document}

\title{An approach to constructing genuinely entangled subspaces of maximal dimension}

\author{Maciej Demianowicz and Remigiusz Augusiak}

\institute{Maciej Demianowicz \at
Atomic Physics Division, Department of Atomic, Molecular and Optical Physics, Faculty of Applied Physics and Mathematics, Gda\'nsk University of Technology, Narutowicza 11/12, 80–233 Gda\'nsk, Poland \\
	\email{maciej@mif.pg.gda.pl (corresponding author)}           
	\and
	Remigiusz Augusiak \at
	Center for Theoretical Physics, Polish Academy of Sciences, Aleja Lotnik\'ow 32/46, 02-668 Warsaw, Poland\\
	\email{augusiak@cft.edu.pl}
}

\maketitle

\begin{abstract}
Genuinely entangled subspaces (GESs) are the class of completely entangled subspaces  that  contain only genuinely multiparty entangled states. They constitute a particularly useful notion in the theory of entanglement
but  also have found an application, for instance, in quantum error correction and cryptography. In a recent study [Phys. Rev. A \textbf{98}, 012313 (2018)], we have shown how GESs can be efficiently constructed in any multiparty scenario from the so--called unextendible product bases. The provided subspaces, however, are not of maximal allowable dimensions and our aim here is to put forward an approach to building such.
The  method is illustrated with few examples in small systems. Connections with other mathematical problems, such as spaces of matrices of equal rank and the numerical range, are discussed.
\end{abstract}


\section{Introduction }

Genuinely entangled states are a crucial resource for many  quantum information processing protocols in networks (see, e.g., \cite{GezaMetro2012,Epping-qkd,GME-dense-coding,Ribeiro-2018,Limited-size-2018}). Their exhaustive characterization is thus of vital importance for the success of future quantum technologies and for this reason it has been the subject of intensive, both theoretical (see, e.g., \cite{polacos-gme-local,Klobus-2019,shen-chen-2019,Zhao-2019}) and experimental (see, e.g., \cite{Barreiro-2013,Micuda-2019,Mooney-2019}), studies.

 A particular line of research on entanglement in multipartite systems concerns characterization of subspaces composed only of entangled states. Primarily, these were  completely entangled subspaces (CESs), that is subspaces only  with states that are in any way entangled \cite{upb-bennett,ces-bhat,ces-partha}. Recently, we have witnessed an interest in so--called genuinely entangled subspaces (GESs), i.e., subspaces composed solely of genuinely multiparty entangled (GME) states, or, in other words, void of states displaying any form of separability \cite{upb-to-ges,ent-of-ges,Wang2019-ges,ManikBanik-ges}.
The initial interest in CESs and GESs stemmed from the observation that (mixed) states supported on them are, respectively, entangled and GME. However, entangled subspaces have also been proved useful in quantum error correction \cite{GourWallach,zahra,Ball,felix-arxiv,AME-alsina}  (in particular,  $k$--uniform subspaces \cite{ces-partha}) and, very recently,  their applicability in cryptographic protocols has been recognized \cite{nonlocal-subspaces}.
It is expected that the range of their applications is much wider and they may be a more general resource in protocols where entangled states already serve as such.

One of the main problems in the area is the construction of entangled subspaces, in particular, those of the maximal possible dimensionality. While it is known how to approach it in the case of CESs, the problem remains unsolved in the general case for GESs
 and only suboptimal with this respect constructions have been put forward \cite{upb-to-ges,ent-of-ges}. The aim of the present paper is to fill this gap and propose an approach to constructing maximal GESs. Our strategy is to select those subspaces from the set of CESs which are at the same time GESs. The main tool of our treatment of the problem is the characterization of bipartite CESs given in \cite{optimal} and its application  boils down to finding the form of full rank matrices satisfying a certain finite set of conditions.

The paper is organized as follows. In Section \ref{preliminaria}, we provide the necessary background and the notation. In section \ref{qubits}, we introduce a general method of constructing maximal GESs in qubit systems and discuss its application mainly in the three--partite case. Further, in section \ref{qudits}, we show how the method can be applied in the multiparty setup with parties holding qudits instead and illustrate it with an example. Section \ref{connections} discusses connections of the main problem with the notions of spaces of matrices of equal rank and the restricted numerical range. 
We conclude in section \ref{konkluzje}, where we also point out some potential future research directions and state open problems.

\section{Preliminaries }  \label{preliminaria}

We  begin with an introduction of the  terminology and the notation.

{\it Notation.}
In the paper we focus on finite--dimensional product Hilbert spaces, denoted  $\calH_{d_1,d_2,\dots, d_n}=\cee{d_1}\otimes \cee{d_2}\otimes \cdots \otimes \cee{d_n}$ or $\calH_{d^n}=\cee{d}\otimes \cdots\otimes \cee{d}$. Subsystems are denoted $A_1,A_2,\dots, A_n =: \bf{A}$ in the general multipartite case or $A,B,\ldots$ for smaller systems. For pure states we use the traditional denotations: $\ket{\psi},\ket{\varphi},\ldots$, often adding subscripts corresponding to respective (groups of) parties, e.g., $\ket{\psi}_{ABC}$. We will use the standard basis for all the parties $\{\ket{i}\}_{i=0}^{d}$ and the kets will be written as row vectors.

{\it Entanglement.}
 An $n$--partite pure state $\ket{\psi}_{A_1A_2\dots A_n}$ is said to be {\it fully product} if it can be written as $\ket{\psi}_{A_1A_2\cdots A_n}=\ket{\varphi}_{A_1}\otimes \ket{\phi}_{A_2}\otimes\cdots \ket{\xi}_{A_n}$. Otherwise it is called {\it entangled}. Among entangled states a particularly interesting class is constituted by {\it genuinely multiparty entangled } (GME) states, i.e., those which {\it cannot} be written as $\ket{\psi}_{A_1A_2\cdots A_n}=\ket{\varphi}_{S}\otimes \ket{\phi}_{\bar{S}}$ for any bipartite cut (biaprtition) $S | \bar{S}$, where $S$ is a subset of the parties and $\bar{S}:=\textbf{A}\setminus S$. In other words, a GME state is not {\it biproduct} with respect to any bipartite cut of the parties. A canonical example of a GME state is the famous GHZ state $\ket{GHZ}=1/\sqrt{2}(\ket{00\cdots 0}+\ket{11\cdots 1})$.
A state $\ket{\psi}$ is called $k$-- product if it is of the form
\beqn
\ket{\psi _{{\otimes}^k}}=\ket{\psi_1}_{S_1}\otimes \ket{\psi_2}_{S_2}\otimes \cdots \otimes \ket{\psi_k}_{S_k},
\eeqn%
 where
 $S_1 \cup S_2 \cup \dots \cup S_k=\textbf{A}$ is a $k$--{\it partition}. In the particular case $k=n$, the vector is fully product; when $k=2$ it is biproduct.

 {\it Completely and genuinely entangled subspaces.}
 It is a well--established fact that there exist nontrivial subspaces containing only entangled states, so called {\it completely entangled subspaces} (CESs) \cite{upb-bennett,ces-bhat,ces-partha}. It has been shown that their maximal achievable dimension for  $\calH_{d^n}$ is
 %
 $ D_{\mathrm{max}}^{\mathrm{CES}}=d^n-n d+n-1 
  								= (d^{n-1}+d^{n-2}+\cdots +1-n)(d-1)$.
  %
A characterization of CESs in the bipartite case with a qubit subsystem, i.e., $\calH_{2,m}$, relevant for our purposes, has been given in \cite{optimal}. We present it in Section \ref{qubits} and further extend it in Section \ref{qudits} to the domain of qudits. 

If one additionally imposes the condition that all states in a CES are not only entangled but their entanglement is genuinely multiparty, one then obtains genuinely entangled subspaces (GESs) \cite{upb-to-ges,ent-of-ges} (see also \cite{ces-partha,schmidt-rank}). Since this notion is crucial in the present paper, we  single out their formal definition.

\begin{defi}
A subspace $\calG \subset \calH_{d_1,\dots , d_n}$ is called a genuinely entangled subspace (GES)  of $\calH_{d_1,\dots , d_n}$ if any $\ket{\psi} \in \calG$ is genuinely multiparty entangled (GME).
\end{defi}
To obtain the maximal available dimension of a GES, one needs to consider maximal dimensions of all bipartite CESs and take the smallest among them. 
It is then easy to see that for $\calH_{d^n}$ \cite{schmidt-rank}:
\beqn
D_{\mathrm{max}}^{\mathrm{GES}}=(d^{n-1}-1)(d-1).
\eeqn
Importantly, it is in fact achievable as a set of randomly chosen $D_{\mathrm{max}}^{\mathrm{GES}}$ vectors will typically span a GES. The achievability can also be seen from the construction given in the present paper. We comment on this issue later in the manuscript.

 An example of a two dimensional GES of $\calH_{2^n}$ is given by the span of the already mentioned $GHZ$ state
 and  the $W$ state, $\ket{W}= 1/\sqrt{n}(\ket{00\dots 001}+\ket{00\dots 010}+\cdots +\ket{10\dots 000})$. In Refs. \cite{upb-to-ges,ent-of-ges} we have given few other constructions of GESs working in general multiparty scenarios attaining larger dimensions. In particular, one of these constructions gives a GES of dimension $d^{n-2}(d-1)^2$. Let us recall it here, for simplicity considering $\calH_{3^3}$. Given is the set of vectors ($\alpha \in \ce$):
 %
$ (1,\alpha+\alpha^3,\alpha^2+\alpha^6)\otimes (1,\alpha^3,\alpha^6)\otimes (1,\alpha,\alpha^2).$ 
 %
The subspace orthogonal to the span of these vectors is a twelve--dimensional GES. Choosing a set of twelve linearly independent vectors of the form above, one obtains an example of  a tripartite non--orthogonal unextendbile product basis.

\section{Maximal GES in qubit systems} \label{qubits}
We now turn to the main body of the paper and propose a construction of GESs of maximal dimensionality. As discussed earlier, our strategy is to use a certain characterization of bipartite CESs related to a one {\it vs} many parties cut and select from them those which are GESs at the same time. 

In this section, we consider multiple qubit Hilbert spaces, i.e., $\displaystyle\calH_{2^n}:=(\cee{2})^{\otimes n}$. The mentioned relevant characterization of bipartite maximal CESs with a qubit subsystem  was given in \cite{optimal}. We recall it below.

%
\begin{fakt}\cite{optimal}\label{optimal-full}
 Let $\calV$ be an $(m-1)$--dimensional CES of $\mathbb{C}^2\otimes \mathbb{C}^m$. Then there exists a nonsingular transformation $\calA: \cee{m} \rightarrow \cee{m}$, such that the following vectors span $\calV^{\perp}$ ($\alpha \in \ce$)
\beq\label{spanning}
\ket{e(\alpha),f_{\calA}(\alpha)}\equiv(1,\alpha)\otimes  \calA \left( 1,\alpha,\alpha^2,\ldots,\alpha^{m-1}\right).
\eeq
\end{fakt}

In our case, $\cee{m}=(\cee{2})^{\otimes(n-1)}$ and we realize that the dimension of the CES agrees with the maximal possible dimension of a GES in this setup: $2^{n-1}-1$. Our aim is to give a characterization of full rank matrices $\calA$ in (\ref{spanning}) leading to GESs.

Before we move to the detailed discussion, let us sketch a general picture of our approach.
The condition that $\calV$ is a GES is equivalent to saying that it is void of any biproduct vectors, i.e., we require vectors of the form   $\ket{\psi}_S\otimes \ket{\phi}_{\bar{S}}$,  for any bipartition $S|\bar{S}$, not to belong to $\calV$. In other words, there can be no such vectors orthogonal to the subspace spanned by the vectors $\ket{e(\alpha),f_{\calA}(\alpha)}$. In what follows, we strictly formalize the latter condition, which in turn characterizes all $\calA$'s leading to GESs. We will refer to such characterization of GESs as the {\it $\calA$--representation}.

It is useful to realize that the task is non--trivial and not all full rank matrices will do the job. With this aim notice that:
%
$(1,\alpha,\alpha^2,\ldots,\alpha^{2^{n-1}-1})_{A_2\dots A_n}=(1,\alpha^{2^{n-2}})_{A_2}\otimes (1,\alpha^{2^{n-3}})_{A_3}\otimes\cdots\otimes (1,\alpha)_{A_n}$.
%
This implies that candidate matrices $\calA$ cannot be product (this in turn precludes, e.g.,  the simplest choice $\calA=\mathbbm{1}$) as locally on $A_1$ and $A_n$ the subspace spanned by  $\ket{e(\alpha),f_{\calA}(\alpha)}$ is then three-dimensional and there thus exists a vector   in $\calV$ which is product across the cut $A_1 A_n|A_2\dots A_{n-1}$. This is most easily seen for three parties with $\calA=\jedynka$. We then have the vectors spanning $\calV^{\perp}$: $(1,\alpha)_A\otimes (1,\alpha^2)_B\otimes(1,\alpha)_C$. The vectors orthogonal to all these vectors are $\ket{\psi_-}_{AC}\otimes \ket{\gamma}_B$, where $\ket{\psi_-}=1/\sqrt{2}(\ket{01}-\ket{10})$ and $\ket{\gamma}$ is arbitrary.

\subsection{General case: $n$ qubits}\label{general-case}


%
Let $\calV$ be a subspace whose orthocomplement $\calV^{\perp}$ is given by Eq. (\ref{spanning}) with some full rank matrix $\calA: (\cee{2})^{\otimes (n-1)} \rightarrow  (\cee{2})^{\otimes (n-1)}$ acting on $A_2,\dots,A_n$ subsystems. Choose a $S|\bar{S}$ bipartition, $S\cup \bar{S}=\textbf{A}$, with $|S|=k$, $|\bar{S}|=n-k$ ($k \le\floor{n/2}$)  and  consider the following (unnormalized) vectors which are product along  this  cut:
\beqn\label{biprodukty}
\ket{\beta,g}:=\ket{\mathbf{\beta}}_S\otimes \ket{g}_{\bar{S}}= (\beta_0^*,\beta_1^*,\ldots,\beta_{2^k-1}^*)_S \otimes (g_0^*,g_1^*,\ldots,g_{2^{n-k}-1}^*)_{\bar{S}}.
\eeqn
The complex conjugation of the elements is for later convenience.
 We assume the parties are ordered lexicographically within each group and the permutation leading to such order is $\sigma$. For example, for $S=A_2 A_5$ and $\bar{S}=A_1 A_3 A_4$, we have $\sigma(12345)=25134$. One can notice that due to this ordering the permutation actually determines uniquely the bipartition. 

A biproduct vector  (\ref{biprodukty}) belongs to $\calV$ if the following holds:
%
\beq \label{warunek-orto}
\inner{\beta,g}{e(\alpha),f_{\calA}(\alpha)}=0, \quad \forall _{\alpha}.
\eeq
Assuming 
\beqn
\calA=\sum_{\substack{i_2,\dots,i_n=0\\i'_2,\dots,i'_n=0}}^1a_{i_2\dots i_n,i'_2\dots i'_n}\outerp{i_2\dots i_n}{i'_2\dots i'_n}
\eeqn
 and representing the indices of $\beta_i$ and $g_j$ in base--$2$,  condition (\ref{warunek-orto}) then rewrites for all $\alpha$:
\beqn\label{polynomial}
\sum_{\substack{ i_{1},\dots, i_{n}=0\\ j_2,\dots, j_n=0}}^1
\alpha^{ i_1+ j_2 2^{n-2}+\cdots+ j_n 2^0}
\beta _{ i_{\sigma(1)}\dots i_{\sigma(k)}} 
 g _{ i_{\sigma(k+1)}\dots i_{\sigma(n)}}
 a_{ i_2\dots i_n,j_2\dots j_n}
 =0.
\eeqn
The LHS of the above is just a polynomial of degree $2^{n-1}$ in $\alpha$. Since the condition must hold for any $\alpha$, each coefficient of this polynomial must be equal to zero, i.e.,
\beqn\label{coeffs-alfa}
\sum_{\substack{ i_1,\dots, i_n=0\\ j_2,\dots, j_n=0\\  i_1+ j_22^{n-2}\dots j_n2^0 = m }}^1
\beta _{ i_{\sigma(1)}\dots i_{\sigma(k)}} 
 g _{ i_{\sigma(k+1)}\dots i_{\sigma(n)}} 
 a_{ i_2\dots i_n,j_2\dots j_n}
 =0,  
\eeqn
where $ m\in\{0,1,\dots,2^{n-1}\} $.
If we now treat $\beta_i^{*}$'s as parameters, Eq. (\ref{coeffs-alfa}) is a homogeneous system of $2^{n-1}+1$ linear equations on $2^{n-k}$ unknowns $g_i^{*}$ with the principal matrix given by:
\beqn\label{principal-matrix}
&&[\calX_{\sigma}]_{m_1\dots m_{n-k},p}:= \\
&&\hspace{+1cm}\sum_{\substack{ i_1,\dots, i_n=0\\ j_2,\dots, j_n=0\\  i_1+ j_22^{n-2}\dots j_n2^0 = p }}^1
\beta _{ i_{\sigma(1)}\dots i_{\sigma(k)}} 
 a_{ i_2\dots i_n,j_2\dots j_n} \delta _{m_1\dots m_{n-k} , i_{\sigma(k+1)}\dots i_{\sigma(n)}} \nonumber
\eeqn
with $m_l \in \{ 0,1 \}$. 
 The demand for a biproduct vector satisfying (\ref{warunek-orto}) not to exist, i.e., $\calV$ to be a GES, requires that the system only has the trivial solution. This only happens when the principal matrix (\ref{principal-matrix}) of the system is full rank, i.e.,
 \beq 
 r(\calX_{\sigma})=2^{n-k}
 \eeq
  for all $\beta_i$'s not being simultaneously zero.  In other words, there cannot be such values
of $\beta_i$'s for which $r(\calX_{\sigma})<2^{n-k}$. The latter condition can be examined using the minors of order $2^{n-k}$ of  $\calX_{\sigma}$. There are ${2^{n-1}+1} \choose {2^{n-k}}$ such minors being (homogeneous) polynomials in $\beta_i$'s, and the rank deficiency of $\calX_{\sigma}$ would require them to have a common root.

We perform analogous analyses for all bipartitions, which is equivalent to all permutations with properly ordered parties and in consequence all $k\le \floor{n/2}$ bipartitions (except  $A_1|A_2\dots A_n$, which by construction does need to be examined as the subspace is a CES across this cut) as mentioned earlier. We thus arrive at the following.
\begin{thm}\label{general-thm} Let $\calV$ be the subspace of $\calH_{2^n}$ orthogonal to the span of the vectors $\ket{e(\alpha),f_{\calA}(\alpha)}$ (\ref{spanning}). Then,
$\calV$ is a GES of dimension $2^{n-1}-1$ iff matrices $\calX_{\sigma}$'s (\ref{principal-matrix}) for all permutations $\sigma$ are full rank for any values of  $\beta$'s. 
\end{thm}
For a given matrix $\calA$ these conditions can be  checked using the Gr\"obner basis \cite{grobner}. Finding explicit form of a GES is an easy task once we know $\calA$ as one can for example determine the projection onto $\calV^{\perp}$ and then find the orthogonal projection.
However, finding a general characterization of $\calA$ for any $n$ and $d$, or, in other words, characterizing the set of all GESs through their $\calA$--representations, seems a hopeless task due to the complexity of the problem. It appears that all one could hope for are examples of classes of good matrices for particular cases. One can also easily construct necessary conditions by considering particular classes of biproduct states not to be present in a subspace. This will be our approach in further parts of the paper.

We should note that a generic $\calA$ will lead to a GES as generically the sets of polynomials under scrutiny will not have common roots. Nevertheless, a random matrix will not be satisfactory from the practical point of view and in further parts we will be interested in some structured examples of constructions.

In what follows, an $\calA$ matrix for a setup with $n$ parties holding $d$ level subsystems will be denoted by $\calA^{(n,d)}$.

\subsection{Three--qubits case}

Let us illustrate the method with the three--qubits case. In principle, in this case it is possible to solve  the problem fully and characterize all matrices $\calA$ for GESs. However, the characterization one obtains is very complicated and does not offer much insight into the structure of the matrices, which could later serve as a hint for generalizations for more parties. We will thus be  satisfied with an exemplary few parameters class of matrices for GESs and
 an easy closed-form necessary condition for the form of $\calA$.

\subsubsection{General case}

The vectors spanning the subspace orthogonal to a GES are now given by:
\beqn\label{AvsBC}
(1,\alpha)_A\otimes \calA^{(3,2)} (1,\alpha,\alpha^2,\alpha^3)_{BC},\quad \alpha \in \mathbb{C},
\eeqn
with a properly chosen full rank matrix 
\beqn
\calA^{(3,2)}=\sum_{m,n=0}^1 \sum_{\mu,\nu=0}^1 a_{m\mu,n\nu}\ket{m\mu}\bra{n\nu}.
\eeqn
The matrix $\calA^{(3,2)}$ must be constructed in such a way that there are no product, across the cuts $B|AC$ 
and $C|AB$,
 non-zero vectors perpendicular to the subspace  spanned by vectors (\ref{AvsBC}). Let us concentrate on the first case, while for the second one the reasoning goes along the same lines with the only difference that the matrix elements are reshuffled in a certain manner.

Let the vectors product across $B|AC$ be written as [cf. (\ref{biprodukty})]
\beqn\label{produktowy-BvsAC}
(1,\beta^*)_B\otimes (f_{00}^*,f_{01}^*,f_{10}^*,f_{11}^*)_{AC}.
\eeqn
%
%
If there existed such a vector in the subspace under scrutiny,
the following would be true for any value of $\alpha$ [cf. (\ref{polynomial})]:
\beqn
\sum_{k=0}^1 \sum_{\substack{m,\mu \\ n,\nu=0}}^1
\beta^m f_{k\mu}a_{m\mu,n\nu}\alpha^{2n+\nu+k}=0,
\eeqn
which is equivalent to the statement that for every power of $\alpha$ in the above its coefficient equals to zero [cf. (\ref{coeffs-alfa})], i.e.,
\beqn \label{na_f}
\sum_{\substack{k,m,n,\mu,\nu=0\\ 2n+\nu+k = j}}^1 
\beta^m a_{m\mu,n\nu}  f_{k\mu}=0, \quad j=0,1,2,3,4.
\eeqn
For any $\beta$, Eq. (\ref{na_f}) is a system of linear equations with the unknowns $f_{k\mu}$  ($k,\mu=0,1$). Its principal matrix is five-by-four and has the elements [cf. (\ref{principal-matrix})]:
\beqn \label{principal-BvsAC}
[\calX_{B|AC}(\beta)]_{j,k\mu}= \sum_{\substack{m,n,\nu=0\\ 2n+\nu+k = j}}^1 
\beta^m a_{m\mu,n\nu}.
\eeqn
In the more appealing matrix form this reads
\beqn
\calX_{B|AC}(\beta)=
\left(
\begin{array}{cccc}
	& & 0 & 0 \\
	\beta \mathfrak{a}_{2}+\mathfrak{a}_{0} & \beta  \mathfrak{a}_{3}+\mathfrak{a}_{1} & &  \\
	&  & \beta  \mathfrak{a}_{2}+\mathfrak{a}_{0} & \beta  \mathfrak{a}_{3}+\mathfrak{a}_{1} \\
	0 & 0 & &  \\
\end{array}
\right),
\eeqn
where $\mathfrak{a}_i$'s are the rows of $\calA$ written as columns.

System (\ref{na_f}) has a nontrivial solution for $f_{k\mu}$ iff there exists a value of $\beta$ such that $r[\calX_{B|AC}(\beta)]<4$. 

Analogously, for the $AB|C$ cut we consider vectors $(g_{00}^*,g_{01}^*,g_{10}^*,g_{11}^*)_{AB}\otimes (1,\gamma^*)_C$ and obtain the corresponding matrix: 
\beqn\label{principal-CvsAB}
[\calX_{C|AB}(\gamma)]_{j,k\mu}=\sum_{\substack{m,n,\nu=0\\ 2n+\nu+k = j}}^1 
\gamma^m a_{\mu m,n\nu},
\eeqn
or, in the matrix form,
\beqn
\calX_{C|AB}(\beta)=
\left(
\begin{array}{cccc}
	& & 0 & 0 \\
	\beta \mathfrak{a}_{1}+\mathfrak{a}_{0} & \beta  \mathfrak{a}_{3}+\mathfrak{a}_{2} & &  \\
	&  & \beta  \mathfrak{a}_{1}+\mathfrak{a}_{0} & \beta  \mathfrak{a}_{3}+\mathfrak{a}_{2} \\
	0 & 0 & &  \\
\end{array}
\right),
\eeqn
which compared to  (\ref{principal-BvsAC}) simply involves the swap of the second and the third row of $\calA^{(3,2)}$.
Again, if there existed a product vector for this cut, there would be a value of $\gamma$ for which $r[\calX_{C|AB}(\gamma)]<4$.

 The following then provides a necessary and sufficient condition for a matrix $\calA^{(3,2)}$ to correspond to an $\calA$--representation of a GES (cf. theorem \ref{general-thm}).

\begin{fakt}\label{nec}
$\calA^{(3,2)}$ corresponds to a GES of $\calH_{2^3}$ iff the matrices $\calX_{B|AC}(\beta)$ 
(\ref{principal-BvsAC}) and $\calX_{C|AB}(\gamma)$  (\ref{principal-CvsAB}) are rank--four
for any $\beta$ and $\gamma$, respectively.
\end{fakt}

In turn, characterizing all GESs in three qubit systems amounts to determining
the form of  $\calA^{(3,2)}$'s  for which $\calX_{B|AC}(\beta)$ and $\calX_{C|AB}(\gamma)$ are full rank for any values of the parameters. 
Since an $m\times n$ ($m \ge n$) matrix has the rank lower than $n$ iff $m$ distinct $n\times n$ minors are zero, the full rank condition on  each of the matrices tells us that all of its five principal minors  cannot vanish simultaneously for some value of the parameter ($\beta$ or $\gamma$), i.e., they cannot have a common root when treated as polynomials in this parameter. In principle, this can be checked analytically for any given matrix as the minors are now polynomials of degree at most four and the methods of solving such polynomial equations are available.
 Unfortunately, we have not been able to obtain a compact closed form characterization of such $\calA$'s. It is nevertheless possible to obtain a simple necessary condition on these matrices by considering particular biproduct states, namely the ones with $\beta,\gamma=0$  and $\beta,\gamma=\infty$, which correspond to, respectively, $(1,0)_{B,C}$ and $(0,1)_{B,C}$. Imposing now that $r[\calX_{B|AC}(0)]=r[\calX_{B|AC}(\infty)]=r[\calX_{C|AB}(0)]=r[\calX_{C|AB}(\infty)]=4$, we obtain the announced necessary condition.
 
\begin{thm}\label{necessary-A}
Let $A_{ij}$ be the submatrix of $\calA^{(3,2)}$ composed of its $i$--th and  $j$--th rows.
%
%
If the subspace orthogonal to the span of vectors (\ref{spanning}) is a GES then neither of the matrices $A_{01}$, $A_{23}$, $A_{02}$, or $A_{13}$ is of the form:
\beqn\label{formy}
\hspace{-0.4cm}\left( \vec{0}, \vec{b},\vec{c}, -\xi^2\vec{b}+	\xi \vec{c} \right), 	%
%
%
	%
\quad	\left( \vec{a},\vec{b}, \left( \xi_2 -\xi_1^2\right) 		\vec{a}+	\xi_1 \vec{b}, -\xi_1\xi_2 		\vec{a}+	\xi_2 \vec{b} \right),
\eeqn
where $\vec{a},\vec{b},\vec{c}\in \cee{2}$, $\xi,\xi_1,\xi_2\in \ce$.
%
\end{thm}

The proof is moved to Appendix \ref{proof-necessary}.

These forms can be further restricted by considering non--existence of other vectors in a subspace, e.g., $\ket{+}\ket{i}\ket{j}$ and $\ket{-}\ket{i}\ket{j}$, $i,j=0,1$.
One can also reduce the number of parameters at the very beginnning and  consider only matrices of the form: $\calA=\proj{0}\otimes (a\proj{0}+b\proj{1})+\outerp{0}{1}\otimes A_{01}+\outerp{1}{0}\otimes A_{10}+\proj{1}\otimes A_{11}$, with $a,b \ge 0$ and
%
%
%
two-by-two matrices $A_{ij}$. This is due to the fact that we can always write $\calA=\sum_{i,j} \outerp{i}{j}\otimes A_{ij}$ and perform $\jedynka \otimes U (\cdot) \jedynka \otimes  V^{\dagger}$ with $U,V$ stemming from the singular value decomposition of $A_{00}$ (this also applies to other blocks). The unitaries are local operations and do not change entanglement properties of the system. The latter approach, however, does not appear to simplify significantly the problem.

\subsubsection{Fully solved class of $\calA^{(3,2)}$ }

Here we give an exemplary class of matrices, which can be fully solved to give the necessary and sufficient conditions.
Consider the following $\calA$ matrix:
\beqn\label{A-solved}
\calA^{(3,2)}(x)=
\left(
\begin{array}{cccc}
	x\ne 0 & 0 & 0 & 0 \\
	0 & a_{1,1} & a_{1,2} & 0 \\
	0 & a_{2,1} & a_{2,2} & 0 \\
	0 & 0 & 0 & 1 \\
\end{array}
\right)
\eeqn
with $a_{1,1} a_{2,2}-a_{1,2} a_{2,1} \ne 0$ to satisfy the full rank condition. 
Using simple algebra one finds that it gives a GES if and only if the following conditions are fullfilled:
\beqn
a_{i,j}\ne 0, \quad
a_{1,1} a_{2,2}-x  \ne 0, \quad 
a_{1,2} a_{2,1}-x \ne 0.
\eeqn
\subsubsection{Decomposition of a Hilbert space into GESs}

Here we give an example of a decomposition of $\calH_{2^3}$ into three GESs, with two of them obviously being maximal, i.e., of dimension $3$. Such decompositions into orthogonal entangled subspaces are known for CESs \cite{without-spanning} and may be of use in quantum error correction.

To this purpose let us consider a particular matrix from the class considered above, namely $\calA^{(3,2)}(x=2)$.
It is easy to verify that the following set of (not orthonormalized) vectors span the corresponding GES:
$
\ket{\varphi_1}=  \ket{001}- \ket{010}-2\ket{011}-2\ket{110}$, $\ket{\varphi_2}=  \ket{001}- \ket{010}+2\ket{011}-2\ket{101}$, $\ket{\varphi_3}=  \ket{001}+ \ket{010}-\ket{100}$.
Let us call it $GES_1^{(3)}$ with the superscript standing for its dimension.
One immediately notices that in its orthocomplement there are the following two GME states which themselves span a two dimensional GES:
$\ket{\psi_1}=\ket{GHZ}=\ket{000}+\ket{111},\;\;\ket{\psi_2}=\ket{001}+\ket{010}+2\ket{100}$.
Any state orthogonal to $\ket{\psi_1}$, $\ket{\psi_2}$, and $GES_1^{(3)}$ must have the form:
$\ket{\tilde{\psi}_3}=a_0\ket{000}+b_0 \ket{001}-b_0 \ket{010}+d_0\ket{011} +(b_0+d_0)\ket{101}+(b_0-d_0)\ket{110}-a_0\ket{111}$.
We also require that the span of $\ket{\psi_1}$, $\ket{\psi_2}$, and $\ket{\tilde{\psi}_3}$ is also a GES. 
An exemplary (unnormalized) state satisfying these conditions is ($a_0=b_0=0$, $d_0=1$):
$\ket{\psi_3}=\ket{011}+\ket{101}- \ket{110}$.
Let us denote:
$
GES_2^{(3)}=\spann \{ \ket{\psi_1}, \ket{\psi_2},\ket{\psi_3} \}.
$
The remaining $2$--dimensional subspace is: $\calH_2=\spann \{ \ket{000}-\ket{111},
\ket{001}-\ket{010}+\ket{101}+\ket{110} \}$.
It is easy to verify that it is again a GES, call it $GES_3^{(2)}$.
In turn, we have the decomposition:
$
\calH_{2^3}=GES_1^{(3)} \oplus GES_2^{(3)} \oplus GES_3^{(2)}$.

\subsection{Four qubits example}\label{four-qubits-ejemplo}

In case of a larger number of parties the characterization is very difficult due to the number of the bipartite cuts which need to be considered. We thus only give an example of a binary symmetric matrix for a GES.
The matrix reads as follows:
\beqn \label{4-kubity-macierz}
\calA^{(4,2)}=
\left(
\begin{array}{cccccccc}
	0 & 0 & 1 & 0 & 0 & 0 & 1 & 1 \\
	0 & 1 & 1 & 1 & 0 & 0 & 1 & 0 \\
	1 & 1 & 1 & 0 & 1 & 1 & 0 & 1 \\
	0 & 1 & 0 & 0 & 1 & 0 & 0 & 1 \\
	0 & 0 & 1 & 1 & 1 & 1 & 0 & 0 \\
	0 & 0 & 1 & 0 & 1 & 1 & 0 & 1 \\
	1 & 1 & 0 & 0 & 0 & 0 & 1 & 0 \\
	1 & 0 & 1 & 1 & 0 & 1 & 0 & 1 \\
\end{array}
\right).
\eeqn
The spanning vectors for this GES are given in Appendix \ref{four-qubits}.

\section{Maximal GES  in qudit systems} \label{qudits}
We now treat the case of higher dimensional subsystems. We concentrate on the case of equal local dimensions but the result can be easily generalized to any dimensions.
It turns out that a reasoning similar to the one given in Section \ref{qubits} can also be successfully applied here. This is due to the following lemma.

\begin{lem}\label{qudit-ces}
Let $\calV^{\perp}$ be the subspace spanned by the vectors ($\alpha \in \ce$)
\beqn \label{general-ges}
(1,\alpha,\ldots, \alpha^{d-1})_{A_1} \otimes \calA (1,\alpha,\alpha^2,\alpha^3,\ldots,\alpha^{d^{n-1}-1})_{A_2A_3\ldots A_n}  , 
\eeqn
with a full rank matrix $\calA: (\cee{d})^{\otimes (n-1)} \rightarrow (\cee{d})^{\otimes (n-1)}$. Then, the subspace $\calV$ orthogonal to $\calV^{\perp}$ is a CES. In particular, all vectors from $\calV$ are entangled  across the $A_1|A_2\dots A_n$ cut. The dimension of $\calV$ is maximal for the given dimensions and reads:
\beq
 \dim \calV = (d^{n-1}-1)(d-1).
\eeq
\end{lem}
This lemma	follows directly from \cite{ces-partha}, where a construction of CESs without the matrix $\calA$ was put forward. This additional element in the construction allows us to select those CESs which are also GESs of $\calH_{d^n}$. The derivation of the conditions on $\calA$ goes along the same lines as in Section \ref{general-case} and we omit it here as it does not provide any additional insight.

We stress that in this case not all CESs are given through the characterization put forward in Lemma \ref{qudit-ces} and in turn not all GESs may be obtained through this approach. In principle, it could even be the case that none of the GESs is characterized in this way. Nevertheless, a generic matrix will again do the job so we are sure that this is not the case. Clearly, the problem of finding a description of good $\calA$'s gets much more involved here even for the tripartite case as there are no closed--form expression for roots of polynomials of degree larger than four. In the general $(n,d)$ case, it is thus natural to consider necessary conditions for the form of the matrix as discussed earlier but even for the simplest cases they get quite involved  and we only give an exemplary matrix in the qutrit case for three parties in the following subsection.

\subsection{Qutrit example}\label{qutrit-ejemplo}
As an illustration, 
we provide a simple binary matrix giving the $\calA$--representation of a GES in the case of three qutrits:
\beqn
\calA^{(3,3)}= \left(
\begin{array}{ccccccccc}
	1 & 0 & 0 & 0 & 0 & 0 & 0 & 0 & 0 \\
	0 & 0 & 1 & 1 & 0 & 0 & 1 & 0 & 0 \\
	0 & 0 & 0 & 1 & 0 & 1 & 1 & 1 & 0 \\
	0 & 1 & 0 & 0 & 0 & 0 & 1 & 1 & 0 \\
	0 & 0 & 0 & 0 & 1 & 0 & 0 & 0 & 0 \\
	0 & 1 & 0 & 1 & 0 & 0 & 1 & 0 & 0 \\
	0 & 0 & 1 & 1 & 0 & 0 & 0 & 1 & 0 \\
	0 & 1 & 0 & 0 & 0 & 0 & 0 & 1 & 0 \\
	0 & 0 & 0 & 0 & 0 & 0 & 0 & 0 & 1 \\
\end{array}
\right).
\eeqn
The explicit form of the spannig vectors for this GES is given in Appendix \ref{three-qutrits}.

%

\section{Connections with other fields}\label{connections}

In this section, we discuss connections of the problem of constructing GESs to other fields, in particular, spaces of matrices of equal rank \cite{Westwick} (see also \cite{dfs} for an application of the concept in the area of quantum error correction) and the restricted, in particular local/product, numerical range \cite{restricted-range,product-range}. In the former case, the connection is established for the maximal GESs, in the latter -- it is a general relation regardless of the dimension of a subspace.

\subsection{Spaces of matrices of equal rank perspective}
We concentrate here for simplicity on the three qubit case but the argument easily generalizes to other cases as well.

One quickly realizes that the matrix $\calX_{B|AC}(\beta)$ (\ref{principal-BvsAC}) is just of the general form
%
$\calX_{B|AC}(\beta)= \calA_1 + \beta \calA_2$,
%
with five-by-four matrices $\calA_i$ having elements drawn in a certain way from $\calA$. Let us introduce the space of matrices  spanned by $\calA_1$ and $\calA_2$:
%
$\frakX_{B|AC}=\spann \{\calA_1,\calA_2\}$.
%
The condition on $\calA$ to give rise to the $\calA$--representation of a GES, i.e., that  for {\it all} values of $\beta$ it holds:
$r[\calX_{B|AC}(\beta)]=4$,
is then equivalent to the demand that $\frakX_{B|AC}$ is a so--called $4$--subspace, that is, all its elements are rank four. Analogously, one introduces another space of matrices $\frakX_{C|AB}$ stemming from considering biproduct vectors across the $AB|C$ cut. Our problem of finding good $\calA$'s for the three qubit case can be thus phrased as follows.\\
\newline
\noindent \textbf {Problem.} Which full rank $\calA$'s lead to $\frakX_{B|AC}$ and $\frakX_{C|AB}$  being  $4$--subspaces (i.e., containing only rank--$4$ elements )?
\newline

 It should be noted that the connection we have established here is of different nature than the one from \cite{schmidt-rank}, where construction of entangled subspaces was related to the notion of spaces of matrices of bounded (from below) or equal rank.

\subsection{Restricted numerical range perspective}\label{connection-range}

Various notions of a numerical range have appeared in the quantum information literature in the recent years \cite{restricted-range,product-range,higher-rank,c-range,product-higher}. The one relevant for the present problem -- local or product range -- belongs to a general class of the restricted numerical range. Let us recall these notions.

The following set is called the {\it restricted numerical range} of a matrix $A$ \cite{restricted-range}:
\beqn\label{restricted}
\Lambda_{T}(A)=\{ \langle \psi | A    |   \psi \rangle : \|\ket{\psi}\|=1, \ket{\psi}\in \Omega_T      \},
\eeqn
where $T$ specifies the type of pure states. If the states belong to the set of fully product states, denote it $\Omega_{\otimes^n}$ , one then deals with the {\it local} or {\it product numerical range} $\Lambda_{\otimes ^n}$ \cite{product-range}.

We propose to  consider a more general notion, namely that of the $k$--{\it product numerical range} of a matrix $A$, which we define as follows
\beqn
\Lambda_{\otimes^k}(A)=\{ \langle \psi | A    |   \psi \rangle : \|\ket{\psi}\|=1, \ket{\psi}\in \Omega_{\otimes ^k}      \},
\eeqn
where $\Omega_{\otimes ^k}$ is the set of $k$--product vectors.
For $k=n$ this notion is equivalent to the above--defined product numerical range, which we now propose to call {\it fully product} one to avoid confusion. In the particular case of $k=2$, we have the {\it biproduct numerical range} $\Lambda_{\otimes ^2}$. The trivial case $k=1$  simply recovers the {\it numerical range} of $A$, $\Lambda(A)$ \cite{Hausdorff}. 
Obviously, for a given matrix, the following inclusion relation holds:
$
\Lambda_{\otimes ^n} \subseteq \Lambda_{\otimes ^{n-1}} \subseteq \cdots \subseteq \Lambda_{\otimes ^2} \subseteq \Lambda_{\otimes ^1}\equiv \Lambda.
$

Let us now discuss the connection of these notions with the problem of determining whether a subspaces is  completely or genuinely entangled.
Assume a decomposition of the whole Hilbert space $\calH$: $\calH=\calP \oplus \calQ$, where  $\calP$ is a GES or a CES with projection $P$; the projector onto $\calQ$ is $Q$, i.e., $P+Q=\jedynka$.
 Clearly, it holds: 
\beqn\label{prodGen}
&&\bra {\psi _{\otimes^2}} Q \ket{\psi _{\otimes^2}} \ne 0, \;\; \forall \psi _{\otimes^2}\;\;\; \mathrm{(GES)}, \\
&&\bra {\psi _{\otimes^n}} Q \ket{\psi _{\otimes^n}} \ne 0, \;\; \forall \psi _{\otimes^n} \;\; \mathrm{(CES)},
\eeqn
%
or, stating it differently:
\beqn
&&\bra {\psi _{\otimes^2}} P \ket{\psi _{\otimes^2}} \ne 1, \;\;\; \forall \psi _{\otimes^2}\;\; \mathrm{(GES)}, \\
&&\bra {\psi _{\otimes^n}} P \ket{\psi _{\otimes^n}} \ne 1, \;\;  \forall \psi _{\otimes^n} \;\; \mathrm{(CES)}.
\eeqn
In consequence, we have the following fact. 
\begin{fakt}
	Let $\calH=\calP \oplus \calQ$, and let $P$ and $Q$ be projections onto, respectively, $\calP$ and $\calQ$.
	Subspace $\calP$ is a \\
	
	(a) GES iff $1 \notin \Lambda_{\otimes^2} (P)$, or, equivalently, $0 \notin \Lambda_{\otimes^2} (Q)$, \\
	
		(b) CES iff $1 \notin \Lambda_{\otimes^n} (P)$, or, equivalently, $0 \notin \Lambda_{\otimes^n} (Q)$.
\end{fakt}

In some applications it might be convenient to consider a more specified notion. Let $\Omega_{\otimes ^k}^{\mathrm{str.}} $ be the set of
strictly $k$--product vectors, that is $k$--product ones for which none of the local vectors can further be written in a product form. We then define the {\it strictly $k$--product numerical range} of $A$ as follows:
$
\Lambda_{\otimes^k}^{\mathrm{str.}} (A)=\{ \langle \psi | A    |   \psi \rangle : \|\ket{\psi}\|=1, \ket{\psi}\in \Omega_{\otimes ^k} ^{\mathrm{str.}}     \}.
$

Concluding, let us note that the set $\Omega_T$ in (\ref{restricted}) can also be taken to be the set  $\Omega_{k-\mathrm{produc.}}$ of  so--called $k$--producible states, that is states which can be written as a product of at most $k$--partite states. We then arrive at the notion of the $k$--{\it producible numerical range}:
$
\Lambda_{k-\mathrm{produc.}} (A)=\{ \langle \psi | A    |   \psi \rangle : \|\ket{\psi}\|=1, \ket{\psi}\in \Omega_{k-\mathrm{produc.}}     \}.
$
This notion is expected to be useful, e.g., in the study of the entanglement depth \cite{ent-depth-1,ent-depth} in multiuser networks.

\section{Conclusions and outlook}\label{konkluzje}
We have considered the problem of construc\-ting genuinely en\-tan\-gled sub\-spaces (GESs) of the maximal possible dimension in any multipartite setup. The solution we have proposed here relies on a certain characterization of completely entangled subspaces (CESs) and boils down to finding a form of full rank matrices fulfilling some finite set of  conditions. Unfortunately, we have not been able to provide a general form of such matrices for any number of parties holding systems of arbitrary dimensions. Nevertheless, we have proposed how to construct necessary conditions for these matrices and found their explicit form in the three qubit case. We have also provided exemplary matrices for some other small systems. Finally, connections with the notions of spaces of matrices of equal rank and the restricted numerical range have been discussed.

The results of the present paper raise the question about a general construction of the matrix $\calA$ working in any dimensions and number of parties. It seems a very difficult task, but it appears  that some methods from different fields might prove useful with this aim.  It may also be possible that some other approach could more easily provide a general construction of GESs.
In particular, it seems that the most promising one might be  based on the notion of spaces of matrices of bounded rank already successfully applied for completely entangled subspaces. This will be considered elsewhere  [M. Demianowicz and R. Augusiak, in preparation].

\begin{acknowledgements}
R.A. acknowledges the support from
the Foundation for Polish Science through the First TEAM
project (First TEAM/2017-4/31) cofinanced by the European
Union under the European Regional Development Fund.
\end{acknowledgements}

\bibliographystyle{spphys}       
\bibliography{cytacje5}

\appendix
\section{GES from section \ref{four-qubits-ejemplo} (four qubits example)}\label{four-qubits}

The  GES corresponding to matrix (\ref{4-kubity-macierz}) is:
$
\mathrm{GES}(\calA^{(4,2)})=
\spann\{
\ket{0100}+\ket{0110}-\ket{0111}+\ket{1000}-\ket{1111},
\ket{0001}+\ket{0010}-\ket{0100}-\ket{0110}-\ket{1110},
2 \ket{0000}+5\ket{0010}-\ket{0011}-\ket{0100}-5\ket{0101}-4\ket{0110}-\ket{0111}+2 \ket{1100},
2\ket{0000}-\ket{0001}+\ket{0010}+\ket{0011}+2\ket{0100}-3\ket{0101}-\ket{0110}-\ket{1001},
2\ket{0000}-\ket{0001}+2\ket{0010}+\ket{0011}+\ket{0100}-4 \ket{0101}-2\ket{0110}-\ket{1000}+\ket{1101},
\ket{0000}-\ket{0001}-\ket{0010}+2\ket{0011}+ \ket{0100}-2\ket{0101}+\ket{0111}-\ket{1000}+\ket{1011},
4\ket{0000}-4\ket{0001}+\ket{0010}+3\ket{0011}+3\ket{0100}-7\ket{0101}-2\ket{0110}+\ket{0111}-2\ket{1000}+2\ket{1010}
\}.$
%
That this is indeed a GES can be verified with the Gr\"obner basis for the corresponding set of polynomials as given in the general formulation of the method or with the techniques considered in \cite{ent-of-ges}.

\section{GES from section \ref{qutrit-ejemplo} (three qutrits example)}\label{three-qutrits}
The GES is as follows:
$
\mathrm{GES}(\calA^{(3,3)})=\spann\{\ket{010}-\ket{021}-\ket{211},\ket{001}-\ket{010}-\ket{020}+2\ket{021}-\ket{100},\ket{001}+\ket{011}-\ket{012}+\ket{021}-\ket{112},\ket{001}-\ket{012}+\ket{021}+\ket{022}-\ket{110},\ket{001}-\ket{012}-\ket{020}+\ket{021}-\ket{122}+\ket{221},2\ket{001}-\ket{010}-\ket{012}-\ket{020}+2\ket{021}-\ket{200},\ket{001}-\ket{011}-\ket{012}-\ket{020}+\ket{021}-\ket{022}+\ket{120},\ket{001}-2\ket{010}-\ket{011}-\ket{020}+2\ket{021}-\ket{022}+\ket{102},\ket{001}+\ket{002}-2\ket{010}-\ket{020}+2\ket{021}+\ket{022}-\ket{212},\ket{001}-\ket{012}-\ket{020}+\ket{021}-\ket{022}-\ket{122}+\ket{210},2\ket{001}-\ket{010}-\ket{011}-\ket{012}-2\ket{020}
+2\ket{021}+\ket{101},2\ket{001}-\ket{010}-\ket{012}-\ket{020}+2\ket{021}+\ket{022}+\ket{121},2\ket{001}+\ket{002}-2\ket{010}-\ket{012}-2\ket{020}+3\ket{021}-\ket{111},2\ket{001}+\ket{002}-2\ket{010}+\ket{011}-\ket{012}-2\ket{020}+3\ket{021}+\ket{122}-\ket{220},\ket{001}+\ket{002}-\ket{010}-\ket{012}-\ket{020}+2\ket{021}+\ket{022}+\ket{122}-\ket{202},
\ket{001}+\ket{002}-2\ket{010}+\ket{011}-\ket{012}-2\ket{020}+3\ket{021}+\ket{022}-\ket{201}
\}.$

%
\section{Proof of Theorem \ref{necessary-A}. }\label{proof-necessary}

Let $\calA^{(3,2)}=\sum_{m,\mu,n,\nu=0}^1 a_{m\mu,n\nu}\ket{m\mu}\bra{n\nu}$. The superscripts $(3,2)$ will be omitted onwards. The proof operates on the degree--two minors:
%
$m_{ij,kl}= 	a_{\hat{i},\hat{k}} a_{\hat{j},\hat{l}} - a_{\hat{i},\hat{l}}	a_{\hat{j},\hat{k}}$
%
$i,j,k,l=0,1,2,3$, where $\hat{x}$ denotes the binary representation of a number $x$.
%
Conditions for rank deficiency of matrices $\calX_{B|AC}(0)$, $\calX_{B|AC}(\infty)$, $\calX_{C|AB}(0)$, and $\calX_{C|AB}(\infty)$ involve the minors for, respectively,  $ij=01,23,02,13$, and in this sense they are decoupled.  Moreover, all the conditions  have an identical structure for any value of $ij$, meaning that these cases do not need to be treated separately, but rather collectively, and  the obtained characterization must be valid for all $ij$'s in the range. For clarity the minors will thus  shortly be written as  $m_{kl}$ with subscripts denoting columns from which the elements of $\calA$ are drawn.
The said conditions, with the above notation convention, are given by the system of equations:
\beqn\label{uklad-minorow}
\left\lbrace
\begin{array}{c}
	m_{01} \left(m_{03}+m_{12}\right)-m_{02}^2=0, \\
	m_{01} m_{13}-m_{02} m_{03}=0, \\
	m_{01} m_{23}-m_{03}^2=0, \\
	m_{02} m_{23}-m_{03} m_{13}=0, \\
	m_{23}\left(m_{03}+m_{12}\right) -m_{13}^2=0.
\end{array}\right.
\eeqn

Our strategy is to find forms of $\calA$'s, for which either of the above systems has a solution under $\det \calA \ne 0$. Negating them we obtain the forms which $\calA$ cannot assume, i.e., a necessary condition on its form. Already here we notice that the trivial solutions $m_{kl}=0$ for all $kl$ (for any $ij$) are not allowed as they do not comply with the full rank condition on $\calA$.

First, we simply inspect (\ref{uklad-minorow}) without caring for the fact that the variables are the minors of $\calA$ and the matrix must be full rank --- these assumptions will enter the proof only later.
The analysis of (\ref{uklad-minorow}) will be split into two cases:  ($1$) $m_{01}=0$ and ($2$) $m_{01}\ne 0$ within which possible subcases will be analyzed.

\begin{itemize}
	\item[($1$)] The condition $m_{01}=0$ implies $m_{02}=m_{03}=0$. We have further:
	\begin{itemize}
		\item[($1x$)] if $m_{12}=0$ then $m_{13}=0$, while $m_{23}$ is arbitrary,
		\item[($1y$)] if $m_{12}\ne 0$   then  $\displaystyle m_{23}=m_{13}^2/m_{12}$.
	\end{itemize}
	\item[($2$)] For $m_{01}\ne 0$, we have the following subcases:
	\begin{itemize}
		\item[($2x$)] if $m_{02}=0$ then $m_{03}+m_{12}=0$, $m_{13}=0$, and $\displaystyle m_{23}=m_{03}^2/m_{01}$,
		\item[($2y$)] For $m_{02}\ne 0$, we have the following 
		\begin{itemize}
			\item[($2y^{(1)}$)] if $m_{03}=0$ then $\displaystyle m_{12}=m_{02}^2/m_{01}$ and $m_{13}=m_{23}=0$,
			\item[($2y^{(2)}$)] if $m_{03} \ne 0$ then $  m_{12}=(m_{02}^2-m_{01} m_{03})/m_{01}$, $ m_{13}=m_{02} m_{03}/m_{01}$, and $ m_{23}=m_{03}^2/m_{01}$.
		\end{itemize}
	\end{itemize}
\end{itemize}

Let us now take into account that $m$'s are minors of full rank $\calA$ and see what structures of $\calA$ are possible if (\ref{uklad-minorow}) holds.
For any $ij$ let $A_{ij}$ be the two-by-four matrix residing in the $i$--th and the $j$--th row of $\calA$. Let us write this matrix as
\beqn\label{A-wektory-bloki}
A_{ij}=
\left(
	\vec{a}_0 \; \vec{a}_1 \; \vec{a}_2 \; \vec{a}_3 
\right)
\eeqn
with two-dimensional column vectors $\vec{a}_{i}$. Notice that the condition $r(A)=4$ requiers  $r(A_{ij})=2$. 
Using the notation above, we have for the minors:
$m_{kl}=|\vec{a}_k , \vec{a}_l |$,
where $|\vec{x},\vec{y}|$ is the determinant of a two-by-two matrix with columns being $\vec{x}$ and $\vec{y}$.

We now list the consequences of the conditions derived above.
\\
\\
\noindent
($1$) This condition means that either $\vec{a}_0 =\vec{0}$ (other vectors arbitrary) or $\vec{a}_0 \sim \vec{a}_1 \sim \vec{a}_2 \sim \vec{a}_3$ with $\vec{a}_0\ne \vec{0}$.  In the latter case $r(A_{ij})= 1$, which  contradicts the condition $r(A)=4$. This implies that it must hold $\vec{a}_0 =\vec{0}$. Then, we have:
\\
\\
\noindent
($1x$) [assuming $\vec{a}_0=\vec{0}$] Either $\vec{a}_1=\vec{0}$ or $\vec{a}_1 \sim \vec{a}_2 			\sim \vec{a}_3$ ($\vec{a}_1\ne \vec{0}$) holds, in which case it would be that $r(A_{ij})=1$, again a contradiction with $r(A)=4$. We then conclude that  $\vec{a}_1=\vec{0}$ and
		\beq\label{1xk}
	A_{ij}=(\O ^{2\times 2} ; \tilde{A}_{ij}^{2\times 2})
		\eeq
		with $\det \tilde{A}_{ij}\ne 0$, where $\O ^{2\times 2}$ is the two-by-two zero matrix.
		\\
		\\
		\noindent
($1y$) [assuming $\vec{a}_0=\vec{0}$] Now,  $\vec{a}_1$ and $\vec{a}_2$ are linearly independent. We consider two possibilities  within this case.
\\
\\
\noindent
($1y^{(1)}$) If $\vec{a}_3 =\vec{0}$ the matrix $A_{ij}$ necessarily assume the form: 
\beq \label{1yk1}
 A_{ij}=(\vec{0}, 
\tilde{A}_{ij}^{2\times 2}, \vec{0}) 
\eeq
with full rank $\tilde{A}_{ij}$.
\\
\\
\noindent
($1y^{(2)}$) On the other hand, $\vec{a}_3 \ne \vec{0}$ implies
\beq  
A_{ij}=(\vec{0} ; \tilde{A}_{ij}^{2\times 2};\vec{a}_3)  
\eeq
with $\vec{a}_3$ such that $ m_{23}=	m_{13}^2/m_{12}$. Since  $\tilde{A}_{ij}= \left( \vec{a}_1, \vec{a}_2 \right)$ is full rank, $ \vec{a}_1$ and $\vec{a}_2$ span $\mathbb{C}^2$, one can write $\vec{a}_3=\alpha \vec{a}_1 +\beta \vec{a}_2$ for some $\alpha$ and $\beta$. Now, $m_{23}=|\vec{a}_2,\vec{a}_3|= -\alpha m_{12}$ and $m_{13}=|\vec{a}_1,\vec{a}_3|=  \beta m_{12}$, which finally means  				 
\beq\label{z-wektorem-d}
\hspace{1cm}A_{ij} = \left( \vec{0}, \vec{a}_1,\vec{a}_2, -\beta^2\vec{a}_1+	\beta \vec{a}_2 \right).
\eeq
Notice that $\beta=0$ recovers form (\ref{1yk1}) so the latter form does not need to be considered separately.
\\	\\
\noindent
($2$) We have $A_{ij}=\left( \tilde{A}_{ij} ^{2\times 2}, X_{ij}^{2\times 2} \right)$ with full rank $\tilde{A}_{ij} ^{2\times 2} =\left(\vec{a}_0, \vec{a}_1 \right) $ and $X_{ij}=\left( \vec{a}_2 , \vec{a}_3 \right )$ to be determined.
\\
\\
\noindent
($2x$) The vanishing of $m_{02}=|\vec{a}_0 , \vec{a}_2|$, implies $\vec{a}_2 \sim \vec{a}_0 \ne \vec{0} $. 
\\
\\
\noindent
($2x^{(1)}$) If $\vec{a}_2 =\vec{0}$ then  $m_{03}							=m_{12}=m_{23}=0$, which with the condition $m_{13}=0$ gives
			\beq\label{2xk1}
		A_{ij}=\left( \tilde{A}_{ij} ^{2\times 2} ; \O^{2\times 2} \right).
			\eeq
		($2x^{(2)}_k$) If  $\vec{a}_2 \ne \vec{0}$ then $\vec{a}_2 =\alpha 	\vec{a}_0$ for some $\alpha \ne 0$.  This implies that $m_{12}=|\vec{a}_1 , \vec{a}_2 |=-\alpha m_{01}$ and, in turn, $m_{03}=\alpha 	m_{01}$. Since $m_{13}=|\vec{a}_1 , \vec{a}_3 |=0$, we have $\vec{a}_3=\beta 	\vec{a}_1$, which gives  $\displaystyle |\vec{a}_2 , \vec{a}_3 |=|\alpha\vec{a}_0 , \beta \vec{a}_1 |=\alpha\beta m_{01}= m_{23}=m_{03}^2/m_{01}	= \alpha^2 m_{01}$, from which it follows $\alpha=\beta$. 	Concluding this subcase:
		\beq\label{A-aA}
			A_{ij}=\left( \tilde{A}_{ij} ^{2\times 2} ; \alpha \tilde{A}_{ij} ^{2\times 2} 										\right).
			\eeq
		Althoug we have assumed that $\alpha$ is non--zero, the case $\alpha=0$ can be included here as it simply recovers the form (\ref{2xk1}). Actually, we can go even further and include $\alpha = \infty$, as this reproduces the form (\ref{1xk}).
			\\
			\\
			\noindent
		($2y$) The non--vanishing of $ m_{02}=|\vec{a}_0 , \vec{a}_2 |$  implies that  $\vec{a}_0 $ and  $\vec{a}_2$ are both non--zero and they are not proportional to each other (they are linearly independent).
	\\
	\\
	\noindent
			($2y^{(1)}$) The vanishing of $m_{03}= |\vec{a}_0 , \vec{a}_3 |$ implies $\vec{a}_3 \sim \vec{a}_0$ with $\vec{a}_0\ne \vec{0}$. Further,  $m_{13}= |\vec{a}_1 , \vec{a}_3 |=0$ implies $\vec{a}_3 \sim \vec{a}_1$ with $\vec{a}_1 \ne \vec{0}$ (since $m_{01}\ne 0$). Both conditions can only hold if $\vec{a}_3=\vec{0}$. We thus have:
			\beq
			A_{ij}=\left( \tilde{A}_{ij} ^{2\times 2}, \vec{a}_2, \vec{0} \right)
			\eeq
			with $\vec{a}_2$ such that the 		condition  $ m_{12}=m_{02}^2/m_{01}$ holds. By the same argument as in  ($1y^{(2)}$)  above we have
			\beq
		 A_{ij} = \left( \vec{a}_0,\vec{a}_1, -\beta^2 														\vec{a}_0+	\beta \vec{a}_1, \vec{0} \right).
			\eeq
			\\
			\\
			\noindent
			($2y^{(2)}$) We set $\vec{a}_2=\alpha_1 \vec{a}_0+\beta_1 \vec{a}_1$, $\vec{a}_3=\alpha_2 \vec{a}_0+\beta_2 \vec{a}_1$ with non-zero $\beta_i$'s. By the arguments similar to the ones above we get:
			\beq\label{najogolniejsza}
			 A_{ij} = \left( \vec{a}_0,\vec{a}_1, \left( \beta_2 -\beta_1^2\right) 														\vec{a}_0+	\beta_1 \vec{a}_1, -\beta_1\beta_2 														\vec{a}_0+	\beta_2 \vec{a}_1 \right).
			 \eeq
			We notice that for $\beta_2 =0$ this form reproduces the case  $(2y^{(1)})$ and for $\beta_1 =0$ --- the case $(2x^{(2)})$.

Concluding, the forbidden forms of $A_{ij}$, $ij=01,23,02,13$, are  given by  (\ref{z-wektorem-d}) and (\ref{najogolniejsza}) just as claimed.

\end{document}

%% file: komendy_osid.tex
\newcommand{\beq}[0]{\begin{equation}}
\newcommand{\eeq}[0]{\end{equation}}
\newcommand{\bw}[0]{\begin{widetext}}
\newcommand{\ew}[0]{\end{widetext}}
\newcommand{\bc}[0]{\begin{center}}
\newcommand{\ec}[0]{\end{center}}
\newcommand{\bwn}[0]{\begin{widetext}\begin{eqnarray}}
\newcommand{\ewn}[0]{\end{eqnarray}\end{widetext}}
\newcommand{\beqn}[0]{\begin{eqnarray}}
\newcommand{\eeqn}[0]{\end{eqnarray}}

\newcommand{\proj}[1]{|#1\rangle \langle #1|}
\newcommand{\ket}[1]{|#1\rangle}
\newcommand{\bra}[1]{\langle #1 |}

\newcommand{\outerp}[2]{\ket{#1}\! \bra{#2}}
\newcommand{\inner}[2]{ \langle #1 | #2 \rangle}

\newcommand{\jedynka}[0]{\mathbbm{1}}

\newcommand{\spann}[0]{\mathrm{span}}

\def\calA{{\cal A}}
\def\calB{{\cal B}}
\def\calC{{\cal C}}

\def\calG{{\cal G}}
\def\calH{{\cal H}}

\def\calP{{\cal P}}
\def\calQ{{\cal Q}}

\def\calV{{\cal V}}

\def\calX{{\cal X}}

\def\frakX{{\frak X}}